# *Ultra-Short-Z Linear Collider Parameters*

*G. White, V. Yakimenko, SLAC National Accelerator Laboratory, Menlo Park, CA, USA.*

## *Abstract*

Interest in highly-compressed electron beams has been increasing in recent times, driven by the study of non-linear and even non-perturbative aspects of QED [2]. The FACET-II [7] facility at SLAC is currently (at the time of writing) being constructed and has been predicted to be able to deliver unprecedented peak beam intensities (>200 kA). We consider here what might be possible in pushing the bunch length compression to its limits at a future Linear Collider facility based on experience at FACET and ongoing photo-injector designs. We present an alternative electron-electron collision parameter table for ILC and CLIC colliders in which low charge, round beams with very short (<100nm) bunches are collided. The parameters shown present the possiblility to provide identical luminosities to the existing designs but with lower rf power requirements and/or with improved luminosity quality (fraction of luminosity close to energy peak). Achieving these beam parameters requires further R&D on the bunch compression and beam delivery systems associated with the Linear Colliders, which is discussed.

## *Overview*

The physics of beam-beam interactions at a future Linear Collider (LC) have been well studied [1]. Existing LC design parameters use large horizontal:vertical aspect ratios for colliding beam parameters in order to minimize beamstrahlung radiation to mitigate degradation of the luminosity spectrum and minimize background radiation in particle detectors. The same concerns also limit the minimum available bunch length.

Energy loss during collisions is driven by the photon emission probability per "formation time" $c/l_f$, where the formation length $l_f$ is related to the electron Compton wavelength $\lambda_c/2\pi = \hbar/mc$. Following recent work to probe non-linear QED processes [2], here we consider a regime in which the electron bunches are compressed further than previously considered, such that the bunch collision interaction time becomes comparable to the beamstrahlung formation time. In this regime the beamstrahlung is strongly suppressed. This allows a reduction in the horizontal beam size at collision (the collision of round beams). Following this approach, we present an alternate parameter set for LC designs which are capable of delivering equivalent luminosity, but with lower required beam power leading to operational cost savings. Due to the suppressed beamstrahlung, the alternate paramerers also have reduced detector backgrounds and similar or enhanced luminosity within 1% of the peak collision energy which is critical for much of the particle physics program. A further advantage to this configuration is the elimination of the need for crab cavities within the final focus system due to suppression of sensitivity to the bunch crossing angle at the IP. This significantly reduces tuning time, complexity and risk for achieving the design luminosity.

The realization of the parameters detailed here requires R&D effort for the electron source, the compression system and a new final focus system design to achieve the round-beam configuration. Some details of the required system designs are outlined below.

The new parameters are assumed to be for an electron-electron collider configuration- a positron source with small enough longitudinal emittance to serve for this regime does not currently exist and the operation of a damping ring in the requisite high 5D brightness regime has not been considered.

Luminosity calculations were performed using GUINEA-PIG++ (GP) [3] at 250, 500, 1000 and 3000 GeV collision energies and compared to ILC and CLIC reference designs. Parameter tables are shown below. For the 250 and 500 GeV designs, the small-z parameters allow for reduced power operations at similar luminosities and similar or improved luminosities within 1% of the collision energy peaks. Whilst simulations of 1 and 3 TeV parameters indicate the possibility of greater reduction in beam power and improved beam collision quality, GP (or any other code existing at the time of writing) does not contain all the required physics processes to assess all the relevant backgrounds.

## Electron Beam Source

There is a proposed new generation of electron photoinjectors, to be operated at cryogenic temperatures, that leads to an enhancement of the launch field at the photocathode by a factor of two above the current state of the art [4]. Approximately a factor of 4 lower emittance than the current state-of-the-art LCLS photoinjector: 35 nm-rad range at 100 pC charges is predicted. We use this as our source parameters and assume an electron-electron collision case. Currently envisioned ring designs do not have the requisite 5D brightness to consider a positron option. The APS-U design for example has round, transverse emittances similar to the discussed above cryo-cooled gun, but at least an order of magnitude larger than required longitudinal emittance.

## Luminosity and Choice of IP Beta-function

For a given rf acceleration technology, the luminosity for a Linear Collider may be written in terms of the provided beam power:

$$L = \frac{H_D}{4\pi E} \cdot \frac{N}{\sigma_x \sigma_y} \cdot (P_{rf} \eta_{rf}),$$

where, P and η are the provided rf power and efficiency, N the bunch population, $\sigma_{x,y}$ the horizontal and vertical beam size at the IP, E the beam energy and $H_D$ is an enhancement factor due to a pinching of the vertical spot size during electron-positron collisions.

For the short-z parameter set, $H_D \sim 1$. For the same beam power, the short-z parameter luminosity then reduces linearly with the corresponding existing design $H_D$ parameter but increases according to the existing design $\sigma_x$:$\sigma_y$ ratio.

Emittances in an rf photo-injector scale non-trivially with charge, mainly due to three effects: space-charge ($\varepsilon \propto Q^{2/3}$), rf/chromatic aberration ($\varepsilon \propto Q^{4/3}$), thermal emittance from the cathode ($\varepsilon \propto Q^{1/3}$). Without a specific design, we assume here simply a linear scaling of emittance with charge from the injector of 350nm-rad/nC. The luminosity enhancement over an existing design for an equivalent beam power can be written using the above luminosity formula as:

$$L_{sz}/L_0 = \frac{1}{H_{D,0}} \cdot \frac{N_{sz}}{N_0} \cdot \frac{\varepsilon_0 \beta_0 (\sigma_{0,x}/\sigma_{0,y})}{\varepsilon_{sz} \beta_{sz}},$$

where, the *sz* and *0* subscripts denote the small-z and existing LC designs respectively, ε and β are the geometric emittance and minimum (vertical) beta functions at the IP, $σ_{x,y}$ are the rms spot sizes of the reference design at the IP. Given a fixed source emittance scaling with charge, the one remaining parameter to adjust the luminosity enhancement is the IP beta-function.

Using current ILC and CLIC parameter tables, the maximum required value for $β_{sz}$ to achieve a luminosity enhancement >1 is summarized in the table below.

| Collider Design | $β_{sz,max}$ [μm] | $β_{y,0}$ [μm] |
|---|---|---|
| ILC-250 | 278 | 410 |
| ILC-500 | 500 | 480 |
| ILC-1000 | 785 | 250 |
| CLIC-3000 | 238 | 70 |

The collider design column shows either an ILC or CLIC configuration with the indicated center-of-mass energy in GeV.

In all cases, the requirement to focus a small and equal beta-function at the IP requires a significant re-design of the LC final focus system. In the ILC-250 and ILC-500 cases, a smaller final beta-function than used for the existing design is also required to give an enhanced luminosity. This adds further complexity and risk to the final focus system design and needs to be carefully studied.

## *Beamstrahlung*

The beamstrahlung parameter, denoted as ϒ (also often denoted as $χ$ in the context of High Field QED), is a measure of the field strength experienced in the colliding electron's rest frame. In terms of the critical (Schwinger) field ( $B_c = \frac{m^2 c^2}{e\hbar} = 4.42 \times 10^9\ T = \frac{E_c}{c} = 1.32 \times 10^{18}\ V/m$ ) and the mean field strength [6,8]:

$$\Upsilon = \gamma \frac{B}{B_c},$$

where, $γ$ is the Lorentz factor of the radiating particle. Usually ϒ is defined in terms of the B field and $χ$ in terms of the E field. In the rest frame of an individual electron (or positron) bunch, the E and B fields cancel at high energies (to within a factor of $1/γ^2$), whereas the fields of the opposite bunch are additive in the colliding rest frame. It is convenient to define ϒ in terms of an equivalent magnetic field written in terms of the combined radial electric field and azimuthal magnetic field:

$$B = \frac{E_r}{c} + B_\phi.$$

This is equivalent for either electron-electron or electron-positron collisions, where B fields cancel in the first case and E fields add with the opposite being true in the latter case.

For colliding beams with a Gaussian charge density, the average value of the beamstrauhlung parameter can be estimated as [5]:

$$\langle \Upsilon \rangle \approx \frac{5}{6} \frac{N r_e^2 \gamma}{\alpha (\sigma_x + \sigma_y) \sigma_z},$$

where, N is the number of particles per bunch, $r_e$ the classical electron radius, α the fine structure constant, and $σ_{x,y,z}$ the rms horizontal, vertical and longitudinal bunch sizes. The maximum value of ϒ experienced during the collision for Gaussian beams is $\Upsilon_{max} \sim \frac{12\Upsilon}{5}$ but is determined numerically here by the GP simulation. In the regime $\Upsilon \gtrsim 1$ quantum effects become important, leading to coherent pair creation which is included in the GP simulations. In the regime $\alpha \Upsilon^{2/3} > 1$ (i.e. ~> 1600), perturbative QED breaks down [2] and the physics contained within GP is no longer expected to be correct.

## *Luminosity Calculations and Parameter Scans*

We use the beam-beam collision simulation code GUINEA-PIG++ (GP) to calculate expected luminosities for the short-z parameter case and compare with the existing LC design luminosity calculations (also made with GP and cross-checked with another similar code, CAIN[6]). GP also calculates the energy-weighted luminosity spectrum and the various background processes.

The key to having useable collisions with round beams at the IP is for the bunch length to be comparible or smaller than the beamstrahlung "formation length" which limits the radiated power during the bunch collisions. In the following, the beamstrahlung formation length is approximated (for demonstration purposes) by the following formula for the regime $\Upsilon \gg 1$ [2]:

$$l_f = \frac{\epsilon}{\Upsilon^{2/3}} \frac{\hbar c}{(mc^2)^2}$$

The full machine parameters for the different collision energies considered are presented in the parameter table at the end of this document and include numbers taken from the GP simulations. Some justification for the parameter choices are discussed for the different collision energy configurations below.

## *ILC-250*

In Figure 1 below, we look at how the luminosity quality scales with bunch charge and bunch length. Taking an initial design parameter setpoint as 10nm bunch length at 100pC charge and 35 nm-rad normalized emittance, the charge is varied between 10 and 100 nm, with linear scaling of bunch length to keep the final peak current at ~1MA and linear scaling of emittance. The pulse length/machine rep rate is scaled to keep the same design beam power. The IP beta function is set at 250µm (the smallest considered IP beta function design for any published ILC configuration). As can be seen, the e-e- total luminosity stays fairly constant with the scan whilst the fraction of the luminosity contained within 1% of the design center-of-mass energy increases towards smaller bunch length (the bunch length scales faster than beamstrahlung formation length scale). For comparison, the equivalen e-e+ case is shown- there is a slight increase in total luminosity at higher charge due to an increasing $H_D$ parameter.

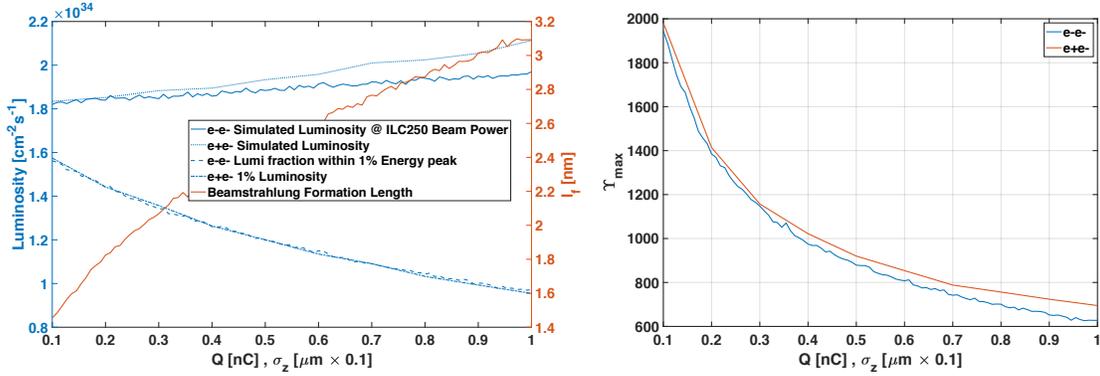

**Figure 1:** (left) Luminosity and beamstrahlung formation length vs. bunch charge (right) beamstrahlung parameter vs. bunch charge. Both emittance and bunch length are scaled linearly from the values corresponding to 100pC bunch charge.

To keep the beamstrahlung parameter in a range where the QED calculations within GP are still likely to be valid ($\Upsilon_{max}$<1600), we choose a bunch charge of 200pC, bunch length of 20nm and emittance of 70 nm-rad for this configuration. The resulting luminosity is 1.87E34 cm$^{-2}$s$^{-1}$ (reference ILC design is 1.61E34) and 1% luminosity is 1.4E34 cm$^{-2}$s$^{-1}$ (reference design is 0.99E34) for the same beam power as the reference design. This would equate to a 14% power reduction over the reference design for the same luminosity, whilst delivering significantly more luminosity within 1% of the peak collision energy (77% vs. 61%).

### ILC-500

To keep $\Upsilon_{max}$<1600 for the 500GeV parameter set we must choose Q>0.6 nC for $\beta^*$=0.48mm (the TDR design value for ILC). If we choose a lower value for the beta-function, the required bunch charge increases along with the luminosity at the expense of degraded collision quality (lower luminosity in the 1% peak). For example, for $\beta^*$=0.35mm, we must choose Q>0.9 nC. Calculated luminosities and beamstrahlung parameters are shown below in Figure 2.

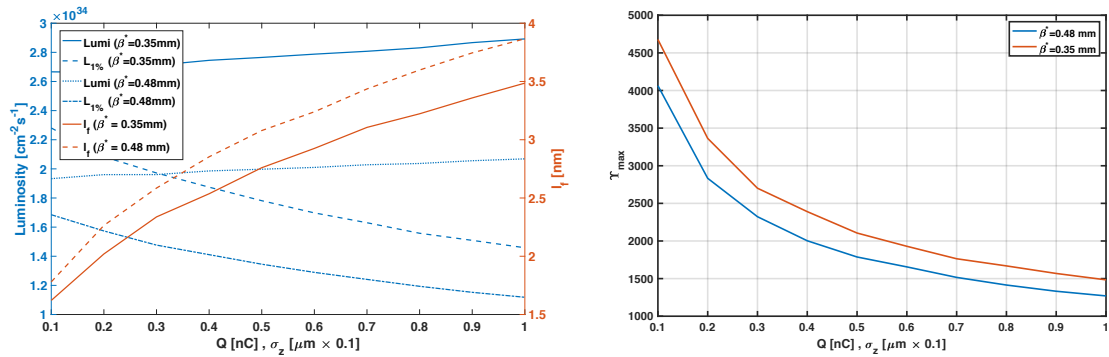

**Figure 2:** (left) Luminosity and beamstrahlung formation length vs. bunch charge (right) beamstrahlung parameter vs. bunch charge. Both emittance and bunch length are scaled linearly from the values corresponding to 100pC bunch charge

For this configuration, we choose Q=0.9 nC, $\beta^*$=0.35mm which gives an enhanced luminosity of 2.9E34 cm$^{-2}$s$^{-1}$ (ILC TDR reference design is 1.8E34). A similar, but slightly lower than TDR design 53% of the luminosity is within the 1% peak (TDR design is 58%).

## ILC-1000

At 1 TeV, there is no region where $Y_{max}<1600$ and the charge is low enough to allow a small bunch length such that beamstrahlung radiation is suppressed. The luminosity and beamstrahlung parameters are shown below for the 1 TeV case for interest, however the GP background calculations are not valid here as QED breaks down in this regime.

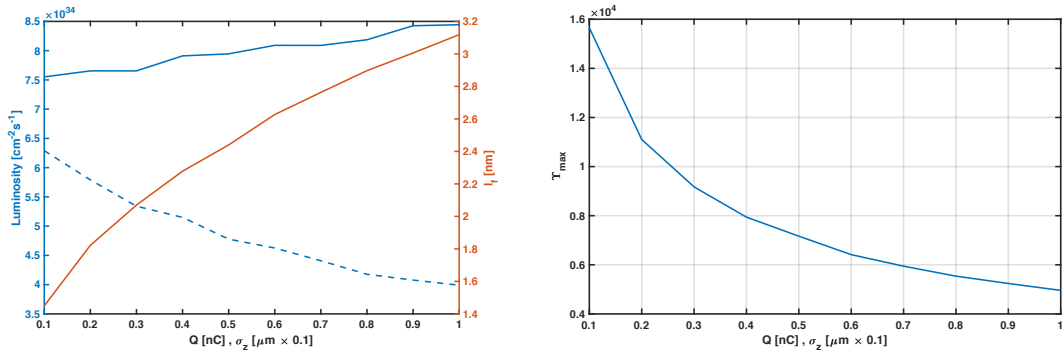

Figure 3: (left) Luminosity and beamstrahlung formation length vs. bunch charge (right) beamstrahlung parameter vs. bunch charge. Both emittance and bunch length are scaled linearly from the values corresponding to 100pC bunch charge

## CLIC-3000

Like the 1 TeV case, there is no valid small-z parameter set for the CLIC 3 TeV collision energy parameters. The luminosities calculated from GUINEA-PIG are much larger than the CLIC design for the same beam power and with lower beamstrahlung energy losses. However, additional non-linear QED processes may change this picture dramatically.

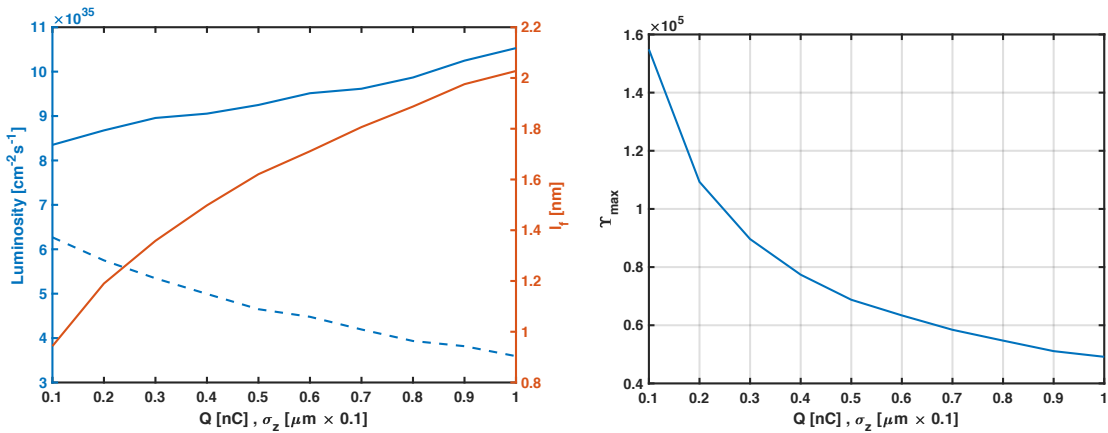

Figure 4: (left) Luminosity and beamstrahlung formation length vs. bunch charge (right) beamstrahlung parameter vs. bunch charge. Both emittance and bunch length are scaled linearly from the values corresponding to 100pC bunch charge

## Design Considerations

The realization of an accelerator capable of delivering the paramers presented here requires considerable R&D. Some of the key challenges requiring further study are summarized below.

## Bunch Compression

The required bunch compression here (to ~1 MA peak current) extends the current state-of-the-art as expressed by the SLAC FACET-II design [7] (1 nC, >200kA) by a factor of 5. Increases in collective effects during bunch compression are expected, and will benefit from recent advances in mitigation strategies, such as CSR compensation and/or shielding techniques [8,9]. Analogous to the FACET-II design, the beam can be compressed from an initial ~0.5 mm to a bunch length of ~10-100 µm using multiple chicane or wiggler-based bunch compression systems and using off-crest rf in the initial Linac stages to generate the required chirp. The limiting factor of the intermediate bunch length may come from tolerable power loads from high-order modes in SC cavities or wakefield effects in high-frequency cavities. The final compression to a bunch length of 10-100nm is expected to be done with a final compression system that is part of the final focus chromaticity correction. Performing the final compression stage helps with CSR emittance degradation, but is a challenge with regards to ISR- the bending systems to perform the compression will have to be weak and therefore long. The number and type of compression stages & compression profile need to be studied and optimized. Consideration of longitudinal space-charge effects during compression is important and the compression profile needs to be well matched to the acceleration profile whilst also taking into account effects such as image-charge heating in the structures with these very high peak current beams.

## Final Focus and Beam Delivery System

The Final Focus System (FFS) can be based on the ILC & CLIC FFS designs, with similar final beta functions and chromaticities. CLIC evaluated 2 designs with similar expected luminosity delivery capabilities: the "traditional" system is similar to the design successfully tested at FFTB at SLAC, whilst the "local chromatic correction" system is similar to the design successfully tested at ATF2, KEK. Whilst being longer, an attractive feature of the traditional optics might be the possibility to include the final compression system into one or both of the chromaticity correction chicanes. Collective effects will need to be carefully considered and the bending systems must be made soft enough to limit ISR growth as with existing BDS bending systems. The energy bandwidth of the FFS is typically <0.5 % which sets the maximum energy spread requirements for the Linac, upstream bunch compression systems and injector.

## Tolerances and Stability

In ILC, beam dynamics studies have shown the bunch compression system to be a major driver for emittance degradation- 6.5 nm out of the total 15nm emittance growth budget is allocated to the bunch compression and ring to main linac transport systems. The main source of emittance growth comes from misalignments and imperfections of the 1.3 GHz accelerating cavities. The rf system also has relatively tight operating tolerances- the tightest tolerance which influences the arrival time of the bunches at the IP is the relative phase of the RF systems on the two sides (0.24° rms of 1.3 GHz). The initial compression stage for this parameter configuration will require 3-30 times more bunch length compression than the nominal design. Further study to understand the implied tolerances on the rf systems and accelerator component alignment requirements are required. Also, the addition of a final bunch compressor stage at high energy will also tighten tolerance requirements and needs further beam dynamics study.

RF and mechanical tolerances required to allow stable collisions may be similar to the existing linear collider designs but needs to be studied. Colliding nm-scale electron-on-electron bunches requires tighter relative collision position tolerance than electron-on-positron, and we also require tight collision tolerances in both transverse planes here. The overall design needs to be studied using start-to-end simulation tools including error sources and dynamic tuning procedures.

For tightly focused beams, the alignment tolerances in the beam delivery system and FFS have been shown to be extremely tight and can only be overcome with active tuning systems. The active tuning systems require precise cancellation of up to 3$^{rd}$ order effects at the IP and themselves have very tight tolerances (e.g. on the resolution of fine-scale magnet mover systems). Experience has shown that the horizontal beta-function at the IP strongly drives the tuning complexity, therefore in this case with a round-beam configuration we should expect an increase in complexity and tolerance requirements in this system which needs careful study.

## *Post-Collision Extraction and Dump*

Beam-beam backgrounds are enhanced for the small-z parameter cases due to copious coherent-pair creation in the large beam-beam fields. Incoherent pair production, on the other hand, is comparatively heavily suppressed. The coherent pairs produced move almost along the direction of the generating photon and mostly don't constitute a background for the detectors from studies at CLIC. Additionally, for the short-z parameter case, contributions of virtual photons to the coherent pair creation are important (so-called trident process). Here they contribute up to 50% of the total pair count. These pairs have an energy spectrum extending to smaller values and can be bent to larger angles in the field of the oncoming bunch and generate detector backgrounds. Previous studies suggest that this background should be negligible but further studies should be carried out here due to the significantly larger number of pairs produced.

The relative energy loss of the colliding beams for the short-z parameters is larger than the corresponding existing e+e- designs. This impacts the chromatic design of the beam extraction system. In the ILC-500 short-z parameter case the post-collision energy spread is outside of the current ILC extraction system design bandwidth and a new design would be required. Although it should be noted that the energy spread is comparable to the CLIC-3 TeV existing design case.

The vertical divergence angle of the beam at the IP in the short-z parameter case is approximately double that of the existing parameter cases. This will significantly reduce (approximately half) the vertical collimation depth. This will increase vertical jitter amplification due to wakefields at the collimator locations and needs to be further studied.

## *Polarization*

The rf photo-injector gun designs assumed here for the initial bunch parameters are not capable of delivering polarized electrons.

*Parameter Table*

| | | ILC-250 | | ILC-500 | | ILC-1000 | | CLIC-3000 | |
|---|---|---|---|---|---|---|---|---|---|
| | | baseline (e+e-) | short-z (e-e-) | baseline (e+e-) | short-z (e-e-) | baseline (e+e-) | short-z (e-e-) | baseline (e+e-) | short-z (e-e-) |
| NOMINAL BUNCH POPULATION | $N$ (x$10^{10}$) | 2.0 | 0.12 | 2.0 | 0.56 | 1.74 | 0.06 | 0.372 | 0.06 |
| SIMULATED LUMINOSITY | $L$ (cm$^{-2}$s$^{-1}$) x $10^{34}$ | 1.6 | | 1.8 | | 3.0 | | 5.9 | |
| LUMINOSITY FRACTION WITHIN 1% | $L_{1\%}$ (%) | 61 | 77 | 58 | 53 | 59 | 83 | 34 | 75 |
| AVERAGE BEAM POWER PER BEAM | $P_b$ (MW) | 2.6 | 2.2 | 5.3 | 3.3 | 10.5 | 4.2 | 14 | 8.2 |
| PULSE FREQUENCY | $f_{rep}$ (Hz) | 5 | 10 | 5 | 10 | 4 | 10 | 50 | 175 |
| BUNCHES PER PULSE | $N_{bunch}$ | 1312 | 9016 | 1312 | 1465 | 2450 | 8354 | 312 | 312 |
| BUNCH SEPARATION | $\Delta t_{bunch}$ (ns) | 554 | 81 | 554 | 496 | 277 | 81 | 0.5 | 0.5 |
| NORMALIZED HORIZONTAL EMITTANCE | $\gamma\epsilon_x$ (μm-rad) | 5.0 | 0.07 | 10.0 | 0.32 | 10.0 | 0.035 | 0.66 | 0.035 |
| NORMALIZED VERTICAL EMITTANCE | $\gamma\epsilon_y$ (μm-rad) | 0.035 | 0.07 | 0.035 | 0.32 | 0.03 | 0.035 | 0.02 | 0.035 |
| HORIZONTAL RMS BEAM SIZE AT IP | $\sigma_x$ (nm) | 515.5 | 8.5 | 474 | 15 | 480.6 | 3.0 | 45.0 | 0.9 |
| VERTICAL RMS BEAM SIZE AT IP | $\sigma_y$ (nm) | 7.66 | 8.5 | 5.9 | 15 | 2.77 | 3.0 | 0.9 | 0.9 |
| LONGITUDINAL RMS BEAM SIZE AT IP | $\sigma_z$ (μm) | 300 | 0.02 | 300 | 0.09 | 250 | 0.01 | 44 | 0.01 |
| HORIZONTAL IP BEAM DIVERGENCE | $\theta_x$ (μrad) | 39.7 | 33.8 | 43 | 42.9 | 21 | 12 | 5.7 | 13 |
| VERTICAL IP BEAM DIVERGENCE | $\theta_y$ (μrad) | 18.7 | 33.8 | 12 | 42.9 | 11 | 12 | 10 | 13 |
| HORIZONTAL IP BETA-FUNCTION | $\beta_x$ (mm) | 13.0 | 0.25 | 11 | 0.35 | 22.6 | 0.25 | 6.9 | 0.07 |
| VERTICAL IP BETA-FUNCTION | $\beta_y$ (mm) | 0.41 | 0.25 | 0.48 | 0.35 | 0.25 | 0.25 | 0.068 | 0.07 |
| BEAMSTRAHLUNG PARAMETER | $Y_{max}$ | 0.067 | 1,388 | 0.146 | 1,570 | 0.305 | 15,632 | 12 | 154,283 |
| VERTICAL DISRUPTION PARAMETER | $D_y$ | 34.5 | 3.9 | 24.6 | 12.7 | 18.7 | 1.9 | -- | 230 |
| LUMINOSITY ENHANCEMENT FACTOR | $H_D$ | 3.1 | 1 | 2.4 | 1 | 1.7 | 1 | 1.4 | 1 |
| # INCOHERENT E+E- PAIRS PER BUNCH CROSSING | $n_{incoh.}$(x$10^6$) | 0.179 | 0.007 | 0.207 | 0.084 | 0.207 | 0.013 | 0.382 | 0.175 |
| # COHERENT E+E- PAIRS PER BUNCH CROSSING *(INC. TRIDENTS)* | $n_{coh.}$ (x$10^6$) | 0 | 212 | 0 | 4,567 | 0 | 15.8 | 941 | 641 |
| AVE. ENERGY LOSS FROM BEAMSTRAHLUNG | $\delta_E$ (%) | 1.9 | 6.4 | 4.5 | 20.6 | 5.6 | 5.8 | 28 | 29 |
| AVE. # PHOTONS / PARTICLE | $n_\gamma$ | 1.62 | 0.43 | 1.72 | 1.50 | 1.97 | 0.35 | 2.1 | 1.3 |